\documentclass[twocolumn,showpacs,amsmath,amssymb,pra,floatfix]{revtex4}

\usepackage{bm} 
\usepackage{epsf} 
\usepackage[english]{babel} 
\usepackage{graphicx} 
\def\au{{\rm \,a.\,u.}} 
\def\pb{$\bar{p}$} 
\def\vmod{$V_{\rm mod}$}%
\def\hmol{{\rm H}$_2$}% 
\def\h2p{H$_{2}^{+}$}% 
\def\hscal{H$_{\rm scal}$}%
 
\def\ket#1{| \, #1 \, \rangle} 
 
\def\opij#1#2#3{\langle \, #1 \, | \, \hat{\rm #2} \, | \, #3 \, \rangle}

\newcommand{\mean}[1]{\left\langle\,#1\,\right\rangle}

\begin{document}

\bibliographystyle{apsrev} 
%\bibliographystyle{unsrt} 

% 
%%%%%%%%%%%%%%%%%%%%%%%%%%%%%%%%%%%%%%%%%%%%%%%%%%%%%%%%%%%%%%%%%%% 
% 
 
\title{A simple parameter-free one-center model potential for an effective
  one-electron description of molecular hydrogen}    
 
\author{Armin L\"uhr} 
\author{Yulian V. Vanne}
\author{Alejandro Saenz} 
 
\affiliation 
{Institut f\"ur Physik,  
AG Moderne Optik, Humboldt-Universit\"at zu Berlin, Hausvogteiplatz 5-7, 
D-10117 Berlin, Germany.}  
%\ead{\mailto{Armin.Luehr@physik.hu-berlin.de}} 

\date{\today} 
 
             
\begin{abstract}\label{txt:abstract} 
For the description of an \hmol\ molecule an effective one-electron model
potential is proposed which is fully determined by the exact
ionization potential of the \hmol\ molecule. In order to test the model
potential and examine its properties it is employed to determine excitation
energies, transition moments, and oscillator strengths in a range of the
internuclear distances, $0.8 < R <  2.5$\au\ \ In addition, it is used as a
description of an \hmol\ target in calculations of the cross sections for
photoionization and for partial excitation in collisions with singly-charged
ions. The comparison of the results obtained with the model potential with
literature data for \hmol\ molecules yields a good agreement and encourages
therefore an extended usage of the potential in various other applications 
%as well as in investigations considering 
or in order to consider
the importance of two-electron and anisotropy effects.  
\end{abstract}

\maketitle

%\clearpage 
%%%%%%%%%%%%%%%%%%%%%%%%%%%%%%%%%%%%%%%%%%%%%%%%%%%%%%%%%%%%%%%%%%%%%%%%%% 
 
% 
\section{Introduction} 
%
%
%
% Motivations: 
%
%
%               Reasons for a simplified treatment of H2 with model potential
%
%  - difficulties to describe molecules fully, because complex system:
%     * two electrons !
%     * anisotropy of electronic charge
%     * orientational dependence of calculated quantities
%     * vibration and rotation of nuclei
%     * potential curves
%    --> complex code needed and significantly more computational power
%
%  - possible comparison to full calculation which include molecular effects
%    due to two electrons, orientation, anisotropy and to determine their 
%    influence on the full results
%    --> how much is H2 like two (modified) H atoms   
%
%
%
%
%
%
%
%
From the very beginning of quantum mechanics the hydrogen atom has been
considered as one of the standard model systems. % of quantum physics.
The reason lies in the simplicity of the theoretical description of this most
basic atomic system. 
On the other hand, the description of the hydrogen molecule is obviously a lot
more involved due to the much larger number of degrees of freedom. Compared to
the atomic case the complexity of the molecule 
arises, e.g., from the electron-electron interaction due to the second
electron and the anisotropy of the charge distribution which may lead to an
orientational dependence of a physical quantity. Additionally, there is
vibrational and rotational motion of the nuclei and 
even in a Born-Oppenheimer approximation one has to deal with  
potential curves for all electronic states and their rovibronic excitations.

Consequently, it would be desirable to have a description, although
simplified,  of the hydrogen molecule at hand which is of similar complexity
as the one of the hydrogen atom. This would allow for an  easy adoption of
already existing numerical methods which were implemented for spherical
one-electron problems to the description of molecular hydrogen. But also in
complex systems including \hmol\ molecules like, e.g., \hmol\ clusters or
\hmol\ adsorbed on surfaces a simple description of the electronic structure
is of interest. 

%In addition, it would reduce efforts in  
%the context of experiments with hydrogen since \hmol\ is experimentally often
%better feasible than H.
A second motivation becomes even more important in the era of fast
improving computational resources which may make the full description
of \hmol\ molecules feasible even in time-dependent processes. That is, the
comparison of results achieved with a full calculation with the outcome of a
simplified description of \hmol\ which has atomic rather than molecular
properties and accounts for the second electron only by screening.  An
analysis of the differences can yield the importance  of the influence of
two-electron as well as of molecular effects, like the deviation
from a spherical symmetric charge distribution.    

In the context of the latter motivation a simple one-electron, single-centered
model potential was proposed in a recent work \cite{sfm:vann08} which deals
with \hmol\ molecules interacting with short intense laser pulses. Since the
strong-field ionization is known to be very sensitive to the electronic
binding energy and the exact form of the long-ranged Coulomb potential the
proposed model potential is designed to agree in these properties with the
\hmol\ molecule. Thereby, the model potential can be adjusted to an arbitrary
internuclear distances by taking the corresponding value of the ionization 
potential. 

Regarding the first motivation, satisfying results have been achieved with the
proposed model potential in the description of an \hmol\ target in
collisions with singly-charged ions \cite{anti:luhr08a}. The calculated total
and differential ionization and excitation cross sections agree well with
literature data down to projectile velocities for which electron-electron
effects may become important. Thereby, also the dependence on the internuclear
distance is examined and the nuclear motion is taken into account. 

The aim of the present work is to further examine the proposed simple
single-centered, effective one-electron model potential and to find out why it
describes the properties of the hydrogen molecule in the applications
\cite{sfm:vann08,anti:luhr08a} to different physical systems so
well. But also the limits of the model in the description of \hmol\
molecules should be analyzed. Therefore, quantities like excitation
energies, electronic transition moments, and oscillator strengths are
calculated as a function of the internuclear distance and are compared to
literature data for a full \hmol\ molecule. 
Also, the model is used to determine photoionization
cross sections and excitation cross sections in collisions with projectiles in
order to test its applicability to different physical systems.
%
%The potential depends
%unambiguously on the exactly known ionization potential of the \hmol\
%molecule which is a function of the internuclear distance. Thereby, the model
%potential also accounts for the internuclear distance $R$ and what is more can
%even deal in some approximation with the motion of the \hmol\ nuclei. 
%
It may be noted, that in the limit $R \rightarrow 0$ the model is also suitable
for the description of atomic helium, as is shortly commented on in the end. 

For a one-electron description of the \hmol\ molecule also
other model potentials exist. To name only three, Teller and
Sahlin~\cite{gen:tell70} discussed a two-center approach while a model
potential for He atoms by Hartree~\cite{gen:hart57} was also
adjusted to \hmol\ by fitting it to the correct ionization potential. It
can be obtained by integrating an effective hydrogen atom-like
charge distribution with Gauss's theorem.  Another 
widely used model is the scaled hydrogen atom model which treats \hmol\ as
a hydrogen atom with a scaled nuclear charge in order to achieve the correct
ionization potential. The latter model is as simple as the one proposed in
\cite{sfm:vann08} but has also the advantage that its wave functions are known
analytically. A disadvantage of the scaled potential is, however, that its
long-range behavior is not correct. It is therefore used in this work for a
comparison of the present results with another \hmol\ model potential.     

% There are ample possible applications for such a model potential. It may be
% used, e.g.,  as a good starting point for the calculation of 
% oscillator strengths as well as for the description of the electronic
% structure of \hmol\  clusters. Satisfying results have already been achieved
% for the description of an \hmol\  target in collisions with singly-charged ions
% \cite{anti:luhr08a} and in interactions with strong laser fields
% \cite{sfm:vann08}. In the former work the nuclear motion was already
% approximately considered. 
% One of the main motivations for the implementation of the model potential may be
% %very valuable for 
% the question %a differential analysis 
% of the origin of features observed in \hmol\ results. 
% For that, \hmol\ results can be compared with those obtained with the
% model potential. An analysis of the differences can yield the
% importance  of the influence of electron-electron effects as well as of
% molecular effects, namely, the deviation of \hmol\ from the spherical
% symmetry.   

The paper is organized as follows: Sec.~\ref{sec:model} presents the model
potential and discusses its properties. Sec.~\ref{sec:applications} considers
various applications of the \hmol\  model, namely, the calculation of excitation
energies, transition moments and oscillator strengths as well as the
determination of cross sections for 
photoionization and excitation in collision processes. Furthermore, the
outcome of these applications is discussed and compared to results of
experiments and theoretical treatments of the full molecular
system. Sec.~\ref{sec:summary} concludes on the findings. 
%and discusses further possible applications of the proposed model. 
Atomic units are used unless otherwise specified.

%%%%%%%%%%%%%%%%%%%%%%%%%%%%%%%%%%%%%%%%%%%%%%%%%%%%%%%%%%%%%%%%%% 
% 
\section{Model potential} 
\label{sec:model} 
% 
%

%
%                Description of (artificial) atomic model
%               ------------------------------------------
%
%  - artificial atom with an isotropic, single-centered charge distribution
%
%  - effective one-electron system, second electron via screening
%
%  - correct electronic binding energies / ionization potential as well as the
%    exact long-range Coulomb potential, since they are very
%    important for ionization and excitation, e.g., in interactions with
%    strong fields or collisions with particles like protons and antiprotons  
%
%  - [solely one parameter alpha which is determined by ionization potential,
%    therefore no additional fit parameter --> simple]
%
%  - no excitation or relaxation of second electron (double excitation) can be
%    described 
%
%  - supposed to be successful as long as two-electron effects play no
%    dominant role
%

In order to obtain a simple model for a complex system 
%one has to make compromises. However, it is important to find 
the right balance has to be found, i.e., a model
which reflects the characteristics of the full description. It is known
that, e.g., ionization processes of \hmol\ are very sensitive to the
ionization potential $I_{{\rm \, H}_2}$ and the properties of the bound states
depend on the exact form of the Coulomb potential. Hence, it is
important that the model potential agrees in these properties with the
molecule. An appropriate trade-off for the description of \hmol\ molecules may
be achieved by using the following simple model potential \cite{sfm:vann08}
for an effective electron with the radial coordinate $r$ 
\begin{equation} 
  \label{eq:model_potential} 
  V_{\rm mod}(r) = - \left(  
                    1 + \frac{\alpha}{|\alpha|}\,\exp\,\left[ 
                                        -\frac{2\,r}{|\alpha|^{1/2}} 
                                      \right ] 
                  \right)\big/\,r,  
\end{equation} 
where $\alpha$ is a dimensionless term. 
% It was proposed by Vanne and Saenz in \cite{sfm:vann08} and further
% discussed in \cite{anti:luhr08a}. They successfully used
% \vmod\ for calculating the ionization and
% excitation of \hmol\   molecules in intense ultrashort laser pulses. A
% comparison with their results from a full molecular treatment of \hmol\ 
% confirmed the 
% applicability of the model for ionization as well as for excitation. 
The model potential satisfies 
the conditions $V_{\rm mod}(r) \rightarrow -1/r$ for $r \rightarrow \infty$ and
describes therefore the long-range behavior of an effective \hmol\ potential
correctly as being hydrogen-atom like.  
Furthermore, it reduces to the potential of a hydrogen atom H for $\alpha
\rightarrow 0$ with an ionization potential $I_{\rm \, H} = 0.5$\au 

\begin{table}[b] 
  \centering 
  \begin{tabular}{@{\hspace{0.2cm}}l@{\hspace{0.5cm}}l@{\hspace{0.5cm}}c@{\hspace{0.2cm}}}
%   \alpha = 1/2 (2 DI + DI^2 + DI^(3/2) Sqrt[4 + DI]),  with DI = I_{H_2} - 0.5
%
%
    \hline 
    \hline 
  $R$ &\multicolumn{1}{l}{\quad$\alpha(R)$}&  $I_{{\rm \, H}_2}(R)$  \\ %\;(\au\,) \\  
    \hline 
%     0.8   &0.348416    &0.341569    &0.715577  \\  %R = 
%     0.9   &0.302668    &0.299135    &0.693373  \\  %R = 
%     1.0   &0.262548    &0.26255     &0.672753  \\  %R = 1.00001
%     1.1   &0.227258    &0.226819    &0.653645  \\  %R = 
%     1.2   &0.196111    &0.1960      &0.635961  \\  %R = 1.20038
%     1.3   &0.168525    &0.1685      &0.619606  \\  %R = 1.30010  
%     1.4   &0.144021    &0.1440      &0.604492  \\  %R = 1.40000 
%                         %-0.607588: Ref. L. Wolniewicz, % J.Chem. Phys. 99
%                         %(1993)
% %    1.4487&0.133081    &0.13308      &0.597555  \\  %R = 1.44871 
%     1.5   &0.122196    &0.1219       &0.590531  \\  %R = 1.50123
% %    1.55  &0.112184    &0.1127       &0.583959  \\  %R = 1.54736
%     1.6   &0.102722    &0.102507     &0.577647  \\  %R = 
% %    1.68  &0.0886449   &0.0881       &0.568062  \\  %R = 1.68326
%     1.7   &0.0853182   &0.0849265    &0.565762  \\  %R = 
%     1.8   &0.0697585   &0.0690       &0.554815  \\  %R = 1.80518
%     1.9   &0.055851    &0.0552638    &0.544745  \\  %R = 
%     2.0   &0.0434376   &0.0434       &0.535499  \\  %R = 2.00001 
%     2.1   &0.0323864   &0.031853     &0.527029  \\  %R = 
% %    2.11  &0.0313519   &0.0313       &0.526223  \\  %R = 2.11088 
%     2.2   &0.0225906   &0.0221641    &0.519292  \\  %R = 
%     2.3   &0.0139698   &0.0136841    &0.512251  \\  %R = 
%     2.4   &0.00646727  &0.00633617   &0.505869  \\  %R = 
%     2.5   &0.000120708 &0.00011624   &0.500115  \\  %R =
% 0.903569884 a.u.  -->  alpha = 0.8791
\vspace{0.075cm}%
    0     &0.87910         &0.903570  \\  %R = 0.0
    0.8   &0.348416        &0.715577  \\  %R = 
    0.9   &0.302668        &0.693373  \\  %R = 
    1.0   &0.262548        &0.672753  \\  %R = 1.00001
    1.1   &0.227258        &0.653645  \\  %R = 
    1.2   &0.196111        &0.635961  \\  %R = 1.20038
    1.3   &0.168525        &0.619606  \\  %R = 1.30010  
    1.4   &0.144021        &0.604492  \\  %R = 1.40000 
                        %-0.607588: Ref. L. Wolniewicz, % J.Chem. Phys. 99
                        %(1993)
    1.4487&0.133081        &0.597555  \\  %R = 1.44871 
    1.5   &0.122196        &0.590531  \\  %R = 1.50123
%    1.55  &0.112184    &0.1127       &0.583959  \\  %R = 1.54736
    1.6   &0.102722        &0.577647  \\  %R = 
%    1.68  &0.0886449   &0.0881       &0.568062  \\  %R = 1.68326
    1.7   &0.0853182       &0.565762  \\  %R = 
    1.8   &0.0697585       &0.554815  \\  %R = 1.80518
    1.9   &0.055851        &0.544745  \\  %R = 
    2.0   &0.0434376       &0.535499  \\  %R = 2.00001 
    2.1   &0.0323864       &0.527029  \\  %R = 
%    2.11  &0.0313519   &0.0313       &0.526223  \\  %R = 2.11088 
    2.2   &0.0225906       &0.519292  \\  %R = 
    2.3   &0.0139698       &0.512251  \\  %R = 
    2.4   &0.00646727      &0.505869  \\  %R = 
    2.5   &0.00012071      &0.500115  \\  %R = 
    \hline 
    \hline 
  \end{tabular} 
  \caption{Values of %the model potential parameter 
    %the dimensionless term 
    $\alpha$ used in this work for different internuclear distances $R$ in
    \au\ \  The ionization potential $I_{{\rm \, H}_2}(R)$ of \hmol\  for these
    $R$ is also given in Hartree.
    It is obtained using the \hmol\  ground-state potential-energy curve
    calculated by Wolniewicz  \cite{dia:woln93}. The ionization potential of a
    He atom \cite{gen:nist08} and the corresponding $\alpha$ value are also
    given as the limit $R \rightarrow 0$.    
    \label{tb:potential-parameter}}  
\end{table} 
% % 
%
The exact dependence of the ionization potential $I_{\rm
  mod}(\alpha)$ on $\alpha$ for a system described by \vmod\ can be determined
numerically (cf.~\cite{sfm:vann08}). However, an advantage of the
model proposed in Eq.~(\ref{eq:model_potential}) is the possibility to
approximate $I_{\rm mod}(\alpha)$ quite accurately with the analytic
expression  
% \cite{sfm:vann08} 
%
\begin{equation}
  \label{eq:ion_potential_approx}
  I_{\rm mod}(\alpha) \approx I_{\rm \, H}
  + {\alpha} \times  \left\{
      \begin{array}{ll}
        {(\,1+\sqrt{|\alpha|}\,)^{{-11}/{4}}}\,, & \alpha < 0\\
        {(\,1+\sqrt{|\alpha|}\,)^{-1}}\,, &        \alpha \ge 0
      \end{array} 
    \right. .
\end{equation}
For instance, at $R=1.4$\au\ the numerically determined ionization potential
and $I_{\rm mod}(\alpha)$ given by Eq.~(\ref{eq:ion_potential_approx}) differ
only by 0.01\%.  The dependence on $\alpha$ simplifies even further in the limit
$|\,\alpha\,| \rightarrow 0$ where the ionization potenial becomes $I_{\rm
  mod}(\alpha) \rightarrow I_{\rm \, H} + {\alpha}$  and depends only
linearly on $\alpha$ as can be seen in table \ref{tb:potential-parameter}.   
 
In order to  describe  an \hmol\ molecule with a fixed
internuclear distance $R$ a certain $\alpha$ has to be determined which
fulfills the requirement that $I_{\rm mod}(\alpha)$ is equal to the ionization
potential $I_{{\rm \, H}_2}(R)$ of the \hmol\  molecule at the considered
fixed distance $R$. In Table \ref{tb:potential-parameter} 
values of $\alpha$ which yield the ionization potentials of \hmol\  for
internuclear distances $R$ in a range from 0.8\au\ to 2.5\au\ are given. For a
fixed $R$ the ionization potential $I_{{\rm \, H}_2}(R)$ is obtained by
subtracting the ground-state potential-energy curve of \hmol\  which was very
accurately calculated by Wolniewicz \cite{dia:woln93} from the ground-state
energies of H$_2^+$. Also given is the $\alpha$ value for the limit $R
\rightarrow 0$ which yields the correct ionization potential of the helium
atom~\cite{gen:nist08}.

Since the model potential can be adopted to different internuclear distances
$R$ with the help of $\alpha$ it is possible to study vibrational effects as
was proposed in \cite{nu:saen97b,dia:erre97}. Ionization cross sections
which account for the vibrational motion of the \hmol\ nuclei in collisions
of \hmol\ targets modeled by \vmod\ with antiprotons were obtained in
\cite{anti:luhr08a}. They were achieved by employing closure,
exploiting the linear behavior of the ionzation cross section with $R$, and
performing the calculations at $R=\mean{R}=1.448$\au\ ($\alpha=0.13308$).

%It should be mentioned that 
However, a molecule treated in the fixed-nuclei
approximation differs from an atom owing to the anisotropy of the electronic
charge distribution which cannot be described correctly within an isotropic,
single-centered atomic model potential. The effect of anisotropy due to both
the two nuclei and due to the second electron in \hmol\  is to some extent
included as a screening of the Coulomb potential. Two-electron effects, like
double excitation or double ionization, are naturally not described properly
by the model.

In order to compare the properties of the proposed model potential \vmod\ in
Eq.~(\ref{eq:model_potential}) with another quite popular (see, e.g.,
\cite{anti:ermo93}) simple artificial atomic model a scaled hydrogen atom
\hscal\ may be introduced. Its potential   
\begin{equation} 
  \label{eq:scaled_potential} 
  V_{\rm scal}(r) = - \frac{Z_{\rm scal}}{r}\,
\end{equation} 
differs from a normal H atom due to the scaled nuclear charge $Z_{\rm
  scal\,}$. The correct ionization potential of \hmol\ at a given $R$ can be
obtained for \hscal , if the nuclear charge is scaled as   
\begin{equation}
  \label{eq:charge_scaling}
  Z_{\rm scal}(R) = \left(\,I_{{\rm \, H}_2}(R)\,
                          \left/\,I_{\rm \, H}\right.\,\right)^{\,1/2}\,. 
\end{equation}
It may be alluded that due to the scaling of the nuclear charge in
Eq.~(\ref{eq:scaled_potential}) all energies $\epsilon_j$ of the bound states
of \hscal\  are affected in the same way, i.e., they are shifted in
comparison to the H atom as   
\begin{equation}
  \label{eq:energy_scaling}
  \epsilon_{j\,}[{\rm H}_{\rm scal}] = (Z_{\rm scal})^2 \;\epsilon_{j\,}[{\rm
    H}]\,. 
\end{equation}
Although the ionization potential of the \hmol\  molecule
is well described by the scaled hydrogen atom it can be expected that this is
not the case for the energies of the bound states, since the potential in
Eq.~(\ref{eq:scaled_potential})  does  
not have the correct $r$ dependence. 
Furthermore, one expects problems in the description of molecular properties
that are very sensitive to the asymptotic long range behavior like tunneling
ionization in intense electric or electromagnetic fields.

% for the target can be obtained if in Eq.\ (\ref{eq:scaled_potential}) since
% the long-range behavior of the
% (\ref{eq:model_potential}) the parameter $\alpha$ is set to $\alpha
% \rightarrow 0$ and the nuclear charge is scaled to $Z_n = Z_{\rm scal} = 1.09$
% as it was, for example, proposed by Ermolaev in \cite{anti:ermo93}. In what
% follows,  it shall be referred to this special case of \vmod\ as $V_{\rm
%   scal}$ which   describes a scaled hydrogen \emph{atom} \hscal\ with
% all energy levels shifted according to
% %
% \begin{equation}
%   \label{eq:energy_scaling}
%   \epsilon_{j\,}[{\rm H}_{\rm scal}] = (Z_{\rm scal})^2 \;\epsilon_{j\,}[{\rm
%     H}]\,. 
% \end{equation}
% %
% The ionization potential of \hscal\ is equal to the absolute value of
% the ground-state energy
% $I_p[{\rm H}_{\rm scal}]=\left|\,\epsilon_{1\,}[{\rm H}_{\rm scal}] \,\right|
% = 0.59405$\au\ and corresponds to the ionization potential of a \hmol\  molecule
% at an internuclear separation $R=1.474$\au\ \   
 
%
The physical quantities studied in this work like oscillator strengths,
transition probabilities, or cross sections obtained with an effective
one-electron model are multiplied with a factor two in order to account for
the two equivalent electrons of \hmol .
%get the quantities for the two-electron system \hmol . 
It should be noted, that also
alternative ways to interprete the results of single-electron models for
two-electron systems
%other interpretations within the independent particle model approach 
are possible (e.g. \cite{sct:wang95}).

% 
%%%%%%%%%%%%%%%%%%%%%%%%%%%%%%%%%%%%%%%%%%%%%%%%%%%%%%%%% 
% 
\section{Applications of the model potential} 
\label{sec:applications} 
% 
% 

%In order to do  such a comparison one has to keep in mind 
%that the model potential describes an artificial atom and consequently,
%that first, no orientational dependence exists and second, 
%its symmetries differ from those of the proper \hmol\  molecule. 
Since the model potential of Eq.~(\ref{eq:model_potential}) is isotropic it is
naturally qualified for describing orientationally-averaged \hmol\ molecules. 
This is often the case in experimental studies in which 
isotropic, non-aligned molecules are investigated.
%no orientational-dependence is induced. 
% The present results obtained with \vmod\ are therefore compared to
% those of an orientationally-averaged \hmol\  which is, however, the normal
% case in experimental studies. 
%
%In the following, it will also be dealt with dipole-allowed transitions from
%the ground state which can be considered as the most prominent ones. In this
%case 
Optical excitations into $p$ states of the model \hmol\  are consequently
compared with both possible dipole-allowed transitions into $^1\Sigma_u$  and
$^1\Pi_u$ states of the \hmol\ molecule. An orientational averaging yields in
this case the factors $1/3$ and $2/3$ for a weighting of the results for the
symmetries $^1\Sigma_u$  and $^1\Pi_u$, respectively. 
On the other hand, the isotropy is of course a limitation of the model. 
For example, in the case of multiphoton excitations interference terms prevent
a determination of simple weighting factors \cite{sfm:apal02,sfm:vann08}.

%By applying the proposed model potential % of Eq.~(\ref{eq:model_potential}) 
%to a number of different purposes it should be investigated  how satisfying
%this model works in practice.
In what follows,  it should be investigated how satisfyingly the proposed model
potential works in practice with respect to various applications.
First, excitation energies, transition moments, and oscillator strengths are
considered. Afterwards, \vmod\ 
is used for the description of ionization and excitation of an \hmol\ molecule
in interactions with photons and in collisions with particles. The findings are
compared to corresponding experimental and theoretical results for an \hmol\
molecule and partly also to results obtained with \hscal.
%
%
%%%%%%%%%%%%%%%%%%%%%%%%%%%%%%%%%%%%%%% 
% 
% 
% 
\subsection{Excitation energies} 
\label{sec:excitation_energies}
\begin{figure}[t] 
    \begin{center} 
      \includegraphics[width=0.46\textwidth]{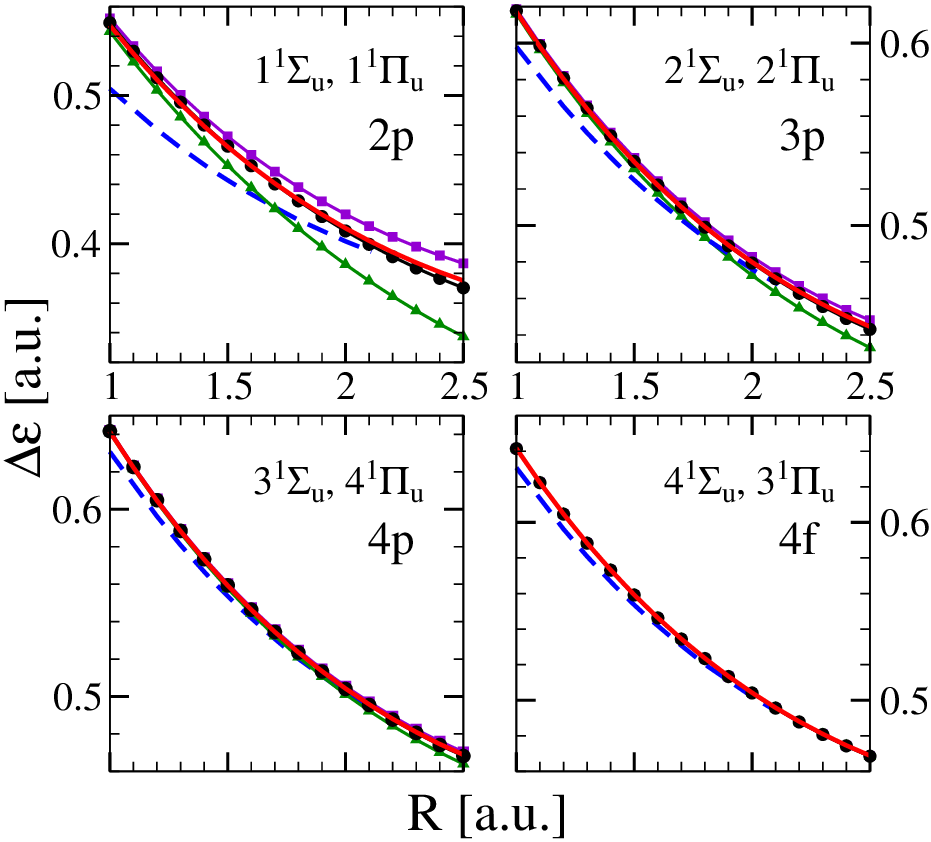} 
      \caption{(Color online) Excitation energies $\Delta\epsilon$ of the
        \hmol\  molecule 
        for the four energetically lowest dipole-allowed final states
        ($n=1,\ldots,4$) for different 
        internuclear distances $R$ calculated by Staszewska and Wolniewicz 
        \cite{dia:stas02,dia:woln03b}: 
        green triangles, $n\,^1\Sigma_u$ states;  
        violet squares, $n\,^1\Pi_u$ states; 
        black circles, orientationally-averaged $^1\Sigma_u$ and $^1\Pi_u$
        (see text). 
        Present excitation energies for corresponding transitions, i.e., from
        the ground state to 2$p$, 3$p$, 4$p$, and 4$f$: 
        red solid curve, model potential; 
        blue dashed curve, hydrogen atom with scaled nuclear charge  \hscal .
        \label{fig:excitation_energies_H2} } 
    \end{center} 
\end{figure} 
In Fig.\ \ref{fig:excitation_energies_H2} excitation energies (EE) for the 
energetically-lowest dipole-allowed final states of the \hmol\  molecule with
the symmetries $n\,^1\Sigma_u$  and  $n\,^1\Pi_u$ with $n=1,\ldots,4$ are
given in the range of internuclear distances $1\au \le R\le 2.5$\au\ \ They
were obtained from the very accurate calculations by Staszewska and
Wolniewicz~\cite{dia:woln93,dia:stas02,dia:woln03b}. The
orientationally-averaged molecular EE are given as circles.  
The corresponding four EE for an atomic system are the energy
differences $\Delta\epsilon$ between the ground state and the 2$p$, 3$p$,
4$p$, and 4$f$ state. 
%The EE for these transitions obtained with the present
%model are shown as solid curves. The corresponding results for \hscal\ are
%indicated by dashed curves. 

It can be seen that in all of the four cases the EE of the
model potential approximates the orientationally-averaged EE
of the \hmol\  molecule very well in the whole $R$ range considered here. Only
for the transition into the 2$p$ state the EE obtained with model potential are
slightly higher than those for \hmol\ for large $R$. It 
is known that in the $R$ range which is considered here the  $4\,^{1}\Sigma_u$
and $3\,^1\Pi_u$ states possess a dominant ($1s4f$) contribution 
%have ($1s4f$) as dominant configuration
\cite{dia:woln03a,dia:spie03}. Consequently, these states cannot be  compared
to a $p$ state of the model potential but instead to the $4f$ state.  

In contrast to the findings for \vmod\ the results for the scaled
hydrogen atom \hscal\ differ from the other $\Delta\epsilon$ curves especially
for small $R$ while they come close to the correct values for $R>2$\au\ \ 
For large $R$ this trend could have been expected since the \hmol\ molecule
becomes more and more like two distant H atoms and therefore can be modeled by
the hydrogen atom-like \hscal. 
%\ atom become more hydrogen atom-like for large $R$. 
However, it is
known that, e.g., the excitation probability can depend considerably on the
EE \cite{anti:luhr08} and therefore should be described
accurately, especially around the equilibrium distance $R\approx1.4$\au

%
%%%%%%%%%%%%%%%%%%%%%%%%%%%%%%%%%%%%%%% 
% 
% 
% 
\subsection{Electronic transition matrix elements} 
\label{sec:transition_elements}
\begin{figure}[t] 
    \begin{center} 
      \includegraphics[width=0.48\textwidth]{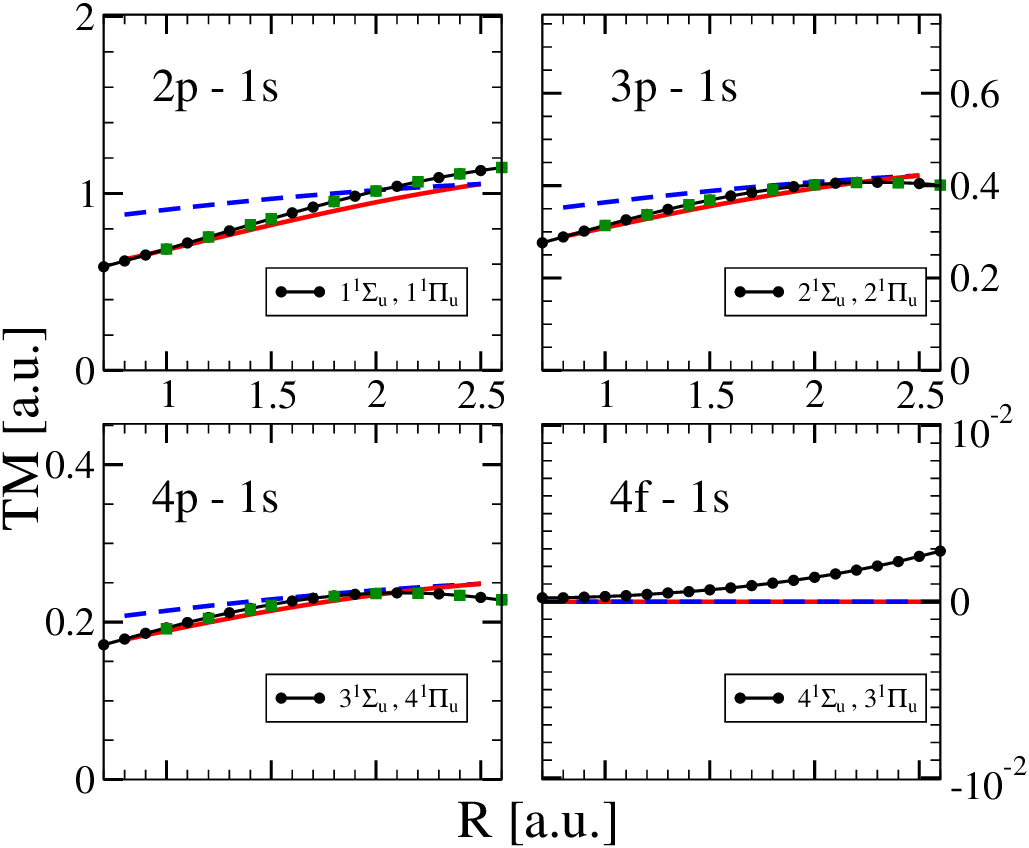} 
      \caption{(Color online) Electronic transition moments (TM) of the 
        \hmol\  molecule as a function of $R$ for transitions 
        from the ground state  $1^1\Sigma_g$ to the four
        orientationally-averaged,  energetically-lowest,
        dipole-allowed final states consisting of the symmetries $^1\Sigma_u$
        and $^1\Pi_u$: 
        black circles, Wolniewicz and Staszewska \cite{dia:woln03a,dia:woln03b};
        green squares, Drira \cite{dia:drir99}. 
        Present results for corresponding transition moments, i.e., from the
        ground state to 2$p$, 3$p$, 4$p$, and 4$f$: 
        red solid curve, model potential \vmod ;
        blue dashed curve, hydrogen atom with scaled nuclear charge
        \hscal. (Note the different scales, especially for the $4f-1s$
        transition.) 
        \label{fig:transition_moments_H2} } 
    \end{center} 
\end{figure} 

Another test of the capability of the model potential \vmod\ given in
Eq.~(\ref{eq:model_potential}) can be performed by considering transition
moments (TM) which are known to be much more sensitive to the behavior of the
wave functions than the energies. The dipole TM into the state $\ket{nl}$ for a
fixed $R$,
\begin{equation}
  \label{eq:transition_moments}
  M(nl) = \sqrt{2}\;\opij{1s}{x}{nl}\,,
\end{equation}
were computed for the same transitions which have been already discussed in
\ref{sec:excitation_energies}, where $n$ and $l$ are the principal and angular
momentum quantum numbers, respectively.  The factor $\sqrt{2}$ in
Eq.~(\ref{eq:transition_moments}) accounts for the two electrons in the \hmol\
molecule. TM from the \hmol\ ground state $1\,^1\Sigma_g$ to the
dipole-allowed final states $^1\Sigma_u$ and $^1\Pi_u$ 
were calculated by Wolniewicz and Staszewska \cite{dia:woln03a,dia:woln03b},
Spielfiedel \cite{dia:spie03}, and Drira \cite{dia:drir99}. In
Fig.~\ref{fig:transition_moments_H2} the orientationally-averaged molecular TM
are compared to the present results obtained with the model potential, whereas
the wrong assignment done in \cite{dia:drir99} for molecular states with
dominant ($1s4p$) or ($1s4f$) configuration is corrected as proposed in
\cite{dia:woln03a,dia:woln03b,dia:spie03}.  Also given are the TM for the
scaled hydrogen atom \hscal .

In general, the present TM achieved with \vmod\ agree with the data for the
full \hmol\ molecule. For $R>1.5$\au\ there is some deviation for the
transition into  the 2$p$ state which could have been expected, since  the
electron-electron 
interaction and the effects due to the two nuclei are most prominent for the
lowest excited states. Otherwise, all TM to higher states match the
literature data very well. It may be noted that even the molecular states
$4\,^1\Sigma_u$ and $3\,^1\Pi_u$ --- both with dominant ($1s4f$) character at
small $R$ --- are again nicely represented by the non-dipole-allowed $4f$
state of the model. The EE as well as the vanishing TM of the $4f$ state match
the corresponding orientationally-averaged results of the \hmol\ molecule.  

The TM calculated for \hscal\ show for all $p$ transitions a different
dependence on $R$ than the TM for \hmol . For small $R$ they are too
large and for $R\rightarrow 2.5$\au\ they approach the TM calculated for \vmod
. The deviations indicate that, especially at small $R$, the properties of the
wave functions obtained with $V_{\rm scal}$ differ considerably from those of
a real \hmol\ molecule.

%
%%%%%%%%%%%%%%%%%%%%%%%%%%%%%%%%%%%%%%% 
% 
% 
% 
\subsection{Oscillator strengths} 
\label{sec:oscillator_strength}
\begin{figure}[t] 
    \begin{center} 
      \includegraphics[width=0.48\textwidth]{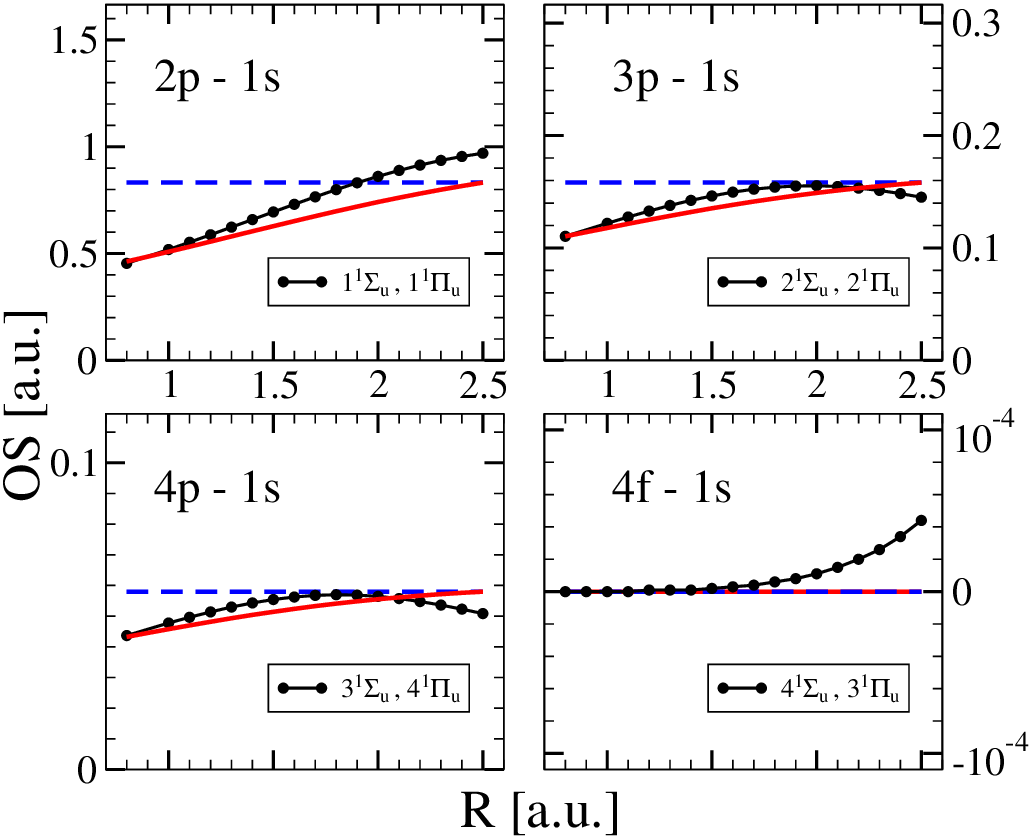} 
      \caption{(Color online) Oscillator strengths (OS) of the   \hmol\
        molecule as a function of $R$ for transitions 
        from the ground state  $1^1\Sigma_g$ to the four
        orientationally-averaged,  energetically-lowest, dipole-allowed final
        states consisting of the symmetries $^1\Sigma_u$ and $^1\Pi_u$: 
        black circles, Wolniewicz and Staszewska
        \cite{dia:stas02,dia:woln03a,dia:woln03b}.
        Results for corresponding transition moments, i.e., from the ground
        state to 2$p$, 3$p$, 4$p$, and 4$f$: 
        red solid curve, present; 
        blue dashed curve,  hydrogen atom with scaled nuclear charge \hscal.
        \label{fig:oscillator_strength_H2} } 
    \end{center} 
\end{figure} 
Fig.~\ref{fig:oscillator_strength_H2} shows the oscillator strengths (OS) of
the \hmol\  molecule, the model potential \vmod , and \hscal\ as a function of
$R$ for the same transitions which were already considered before. It may be
argued that the procedure of orientational averaging is most appropriate for
the OS since they obey the Thomas-Reiche-Kuhn sum rule. Since the OS depend
on the EE and TM the question is whether this leads to a compensation or even
to an increase of the deviations between model and real molecule. The OS 
from the ground into the excited state $\ket{nl}$ are given by 
\begin{equation}
  \label{eq:oscillator_strength}
  f(nl) = \frac{2}{3}\,(\epsilon_{nl}-\epsilon_0)\, |\,M(nl)\,|^{\,2}\, ,
\end{equation}
where $\epsilon_0$ and $\epsilon_{nl}$ are the energies of the ground and
final excited state $\ket{nl}$, respectively. The OS for the \hmol\ molecule
are constructed in the same way using the data calculated by Wolniewicz and
Staszewska \cite{dia:stas02,dia:woln03a,dia:woln03b}. First, the OS for both
symmetries $^1\Sigma_u$ and $^1\Pi_u$ were determined separately and afterwards
orientationally weighted with factors 1/3 and 2/3, respectively, in order to
compare to the present results. 

It can be seen in Fig.~\ref{fig:oscillator_strength_H2} that the OS of \hscal\
are independent of $R$ and are therefore the same as for the hydrogen atom. This
is due to a cancellation  of the $Z_{\rm scal}$ dependence in
Eq.~(\ref{eq:oscillator_strength}). Therein the dependence on  $Z_{\rm  scal}$
of the energies is $\epsilon \propto (Z_{\rm scal})^2$ (cf.\
Eq.~(\ref{eq:energy_scaling})) and of the TM is $M \propto 1/(Z_{\rm scal})$. 
It is even necessary that the OS of \hscal\ are independent of $R$ since
a scaling of the OS with a single factor would lead to a violation of the above
mentioned sum rule.  

The OS of the \hmol\    
molecule and for \vmod\ are, however, not independent of the internuclear
distance $R$. For all transitions the OS of \hmol\ and the present model agree
well for small $R$. For increasing $R$ the OS obtained with \vmod\ increase
roughly linearly while those for \hmol\ show a different behavior for
$R>1.5$\au\ \ However, the magnitudes are still comparable. At $R=2.5$\au\ the
OS obtained with \vmod\ and $V_{\rm scal}$ coincide which already could have
been expected before from the results for the related EE and TM. For this
distance both potentials become hydrogen-atom like. Considering 
the region around $R=1.4$\au\ --- which is most important for many
calculations with fixed $R$ considering processes starting from the \hmol\
ground state --- one can conclude that the OS of the \hmol\
molecule are satisfyingly modeled by the proposed potential \vmod .

%
%%%%%%%%%%%%%%%%%%%%%%%%%%%%%%%%%%%%%%% 
% 
% 
% 
\subsection{Photoionization cross sections} 
\begin{figure}[t] 
    \begin{center} 
      \includegraphics[width=0.475\textwidth]{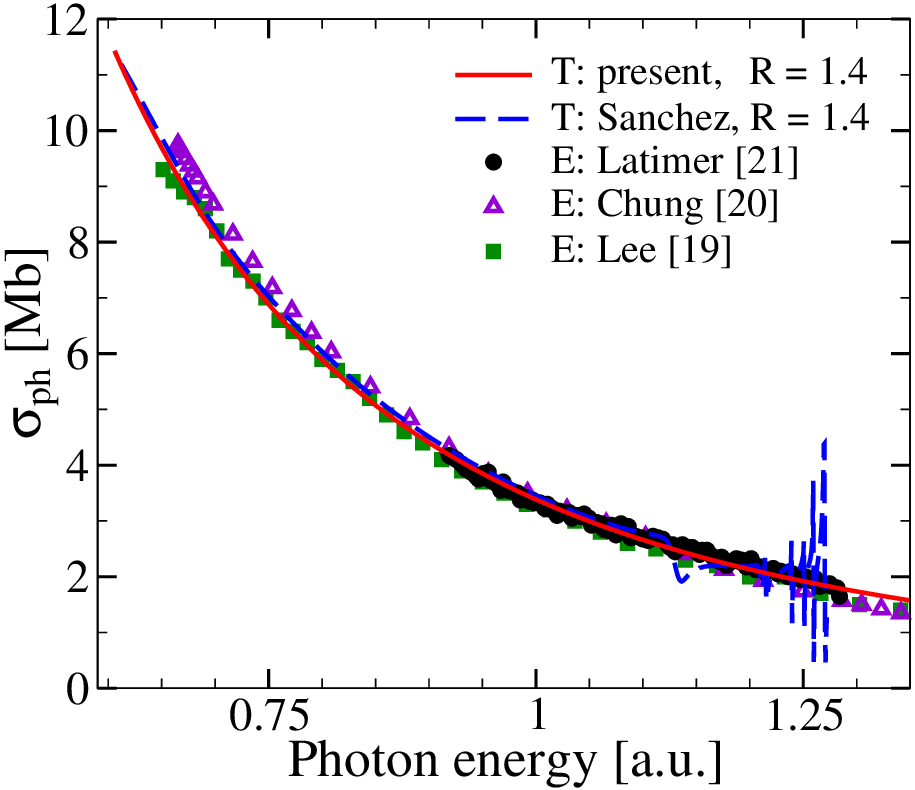} 
      \caption{(Color online) Total photoionization cross section of \hmol\
        as a function of 
        the photon energy for a fixed internuclear distance $R=1.4$\au\ \
        Theory: 
        red solid curve, present results with model potential; 
        blue dashed curve, S{\'a}nchez and  Mart{\'\i}n~\cite{dia:sanc97a}. 
        Experiment: 
        green squares, Lee {\it et al.}~\cite{dia:lee76};
        violet triangles, Chung {\it et al.}~\cite{dia:chun93};
        black circles, Latimer {\it et al.}~\cite{dia:lati95}. 
        \label{fig:cs_photoion_H2} } 
    \end{center} 
\end{figure} 
A calculation of the photoionization spectrum for the hydrogen molecule is used
to demonstrate the applicability of the present model to interaction processes
in which an \hmol\ molecule is ionized. In doing so, the representation of 
the continuum states is probed.  Further applications of \vmod\
in order to describe ionization of \hmol\ in time-dependent processes can be
found elsewhere \cite{anti:luhr08a,sfm:vann08}. The photoionization cross
section is given by
\begin{equation}
  \label{eq:photoionization}
  \sigma_{\rm ph}(\epsilon) = \frac{4\,\pi^2}{c}\,(\epsilon-\epsilon_0)\,
                              \left|\,M(\epsilon)\, \right|^{\,2}\,
                              \rho(\epsilon)\,,
\end{equation}
where $\epsilon$ is the positive energy of the ionized final state
$\ket{\epsilon}$ with an angular momentum $l=1$ and $c$ is the speed of light. 
The transition matrix elements $M(\epsilon)$ are defined in the same way as in
Eq.~(\ref{eq:transition_moments}) except that the $\ket{\epsilon}$ are
considered as final states. The density of continuum states
$\rho(\epsilon)$ is used for energy-normalization of the cross section.

The present photoionization cross sections %obtained with \vmod\ of the model
                                %potential  
were calculated for $R=1.4$\au\ in order to compare the results with
theoretical calculations from  literature in which the fixed-nuclei
approximation was used. 
%which are shown 
Besides the theoretical results by S{\'a}nchez and
Mart{\'\i}n~\cite{dia:sanc97a} also experimental photoionization cross
sections are shown in Fig.~\ref{fig:cs_photoion_H2} which were measured by Lee 
{\it et al.}~\cite{dia:lee76}, Chung {\it et al.}~\cite{dia:chun93} and
Latimer {\it et al.}~\cite{dia:lati95}.

It can be seen that the experimental photoionization cross sections are well
described by the present model. At low energies, however, the results by
Chung~{\it et al.} and Lee {\it et al.} lie slightly above and below the
present curve, respectively. The measurements by Latimer~{\it et al.} where
performed between approximately 0.9 and 1.3\au\ on a dense energy grid
searching for resonances above 1.1\au\ which they did not find. Their data
match very well with the present curve which is, of course, free of any
resonance caused by doubly-excited states. The clearly
visible resonances in the theoretical data calculated by  S{\'a}nchez and
Mart{\'\i}n around $R=1.12$ and 1.25\au\ were explained by Mart{\'\i}n
in~\cite{dia:mart99} as being only visible within the fixed-nuclei
approximation.  In a further calculation which includes nuclear
motion~\cite{dia:mart99} the resonance effects are, in accordance with
experimental results, practically invisible.
This was explained by the broadening of the resonances, if the nuclear degrees
of freedom are included. For energies below 1.1 \au\ where no resonances occur
in the data of~\cite{dia:sanc97a} their calculation
%by S{\'a}nchez and  Mart{\'\i}n 
agrees well with the present curve.

%
%%%%%%%%%%%%%%%%%%%%%%%%%%%%%%%%%%%%%%% 
% 
% 
% 
\subsection{Collisional excitation cross sections} 
\label{sec:ex_cross_section} 
\begin{figure}[t] 
    \begin{center} 
      \includegraphics[width=0.48\textwidth]{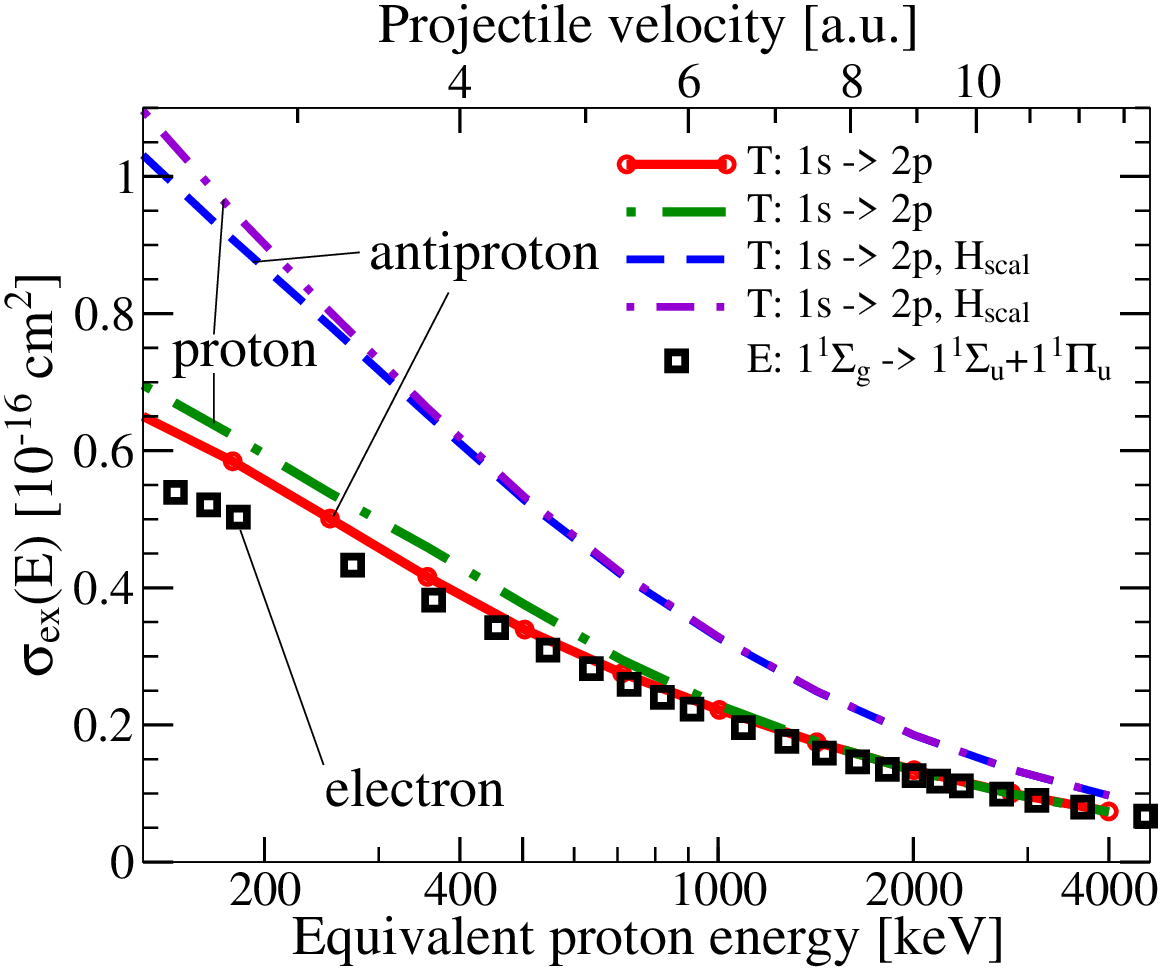} 
      \caption{(Color online) Differential cross sections $\sigma_{\rm ex}$
        for excitation into the lowest dipole-allowed states of \hmol\ as a
        function of the projectile velocity $v$ in \au\ and of the equivalent
        proton energy $E$ in keV.   
        Theory: Present results for a fixed internuclear distance $R=1.4$\au\
        for excitation into 2$p$. Model potential \vmod : 
        red solid curve, antiprotons; 
        green dash-dotted curve, protons.  
        \hscal : 
        blue dashed curve, antiprotons; 
        violet dash-double-dotted curve, protons.  
        Experiment: Sum of cross sections for excitations into $1^1\Sigma_u$ and
        $1^1\Pi_u$: 
        black squares, electrons, Liu {\it et al.}~\cite{sct:liu98}. 
%        For comparison, the impact energies of the electrons are scaled with
%        the ratio of proton to electron mass.  
        \label{fig:cs_ex_H2} } 
    \end{center} 
\end{figure} 

While the ionization probability depends strongly on the ionization potential
%which is by construction correct in the model potential, 
the excitation
process is more sensitive to bound-state properties. Therefore, a calculation
of an excitation cross section for the \hmol\ molecule is used to demonstrate
that the model potential is also capable to describe transitions to bound states
properly.  

In Fig.~\ref{fig:cs_ex_H2} the partial cross section for the
energetically-lowest, dipole-allowed transition for \hmol\ collisions with
protons and antiprotons is shown where the \hmol\ target is described with the
model potential. Detailed information concerning the employed time-dependent
method is given elsewhere~\cite{anti:luhr08a}.  This transition has been
chosen since first, it is the most probable excitation in this energy range
and second, in Sec.~\ref{sec:transition_elements} and
\ref{sec:oscillator_strength} the largest deviation of 
the TM and OS between the model and the \hmol\ molecule have been
observed for this transition. Furthermore, partial cross sections can be used
for a more sensitive testing because the errors of total cross sections may be
reduced by some error compensation. The present results are compared with
experimental data for 
\hmol\  collisions with electrons measured at a temperature of 10\,K by Liu
{\it et  al.}~\cite{sct:liu98} due to the fact that to the best of the authors'
knowledge no measurements have been performed for proton or antiproton impact. 
%The impact energies of the electrons are scaled with the ratio of proton to
%electron mass ($\approx 1822.89$) in order to compare the results in a single
%graph. 
In addition, also results for proton and antiproton collisions with \hscal\ are
given in Fig.~\ref{fig:cs_ex_H2}. 

The results for protons and antiprotons obtained with the proposed model
potential \vmod\ coincide for large impact energies $E>1000$\,keV as 
is expected. At these high energies they also fully agree 
% as predicted by the first Born approximation 
with the experimental data for electrons with the same impact velocity $v$. 
This behavior at high impact velocities is predicted by the first Born
approximation, i.e., the same cross section can be expected for collisions
including particles with the same absolute value of the charge and the same
impact velocity.  
At smaller
energies the cross sections start to depend on the properties of the
projectile. Thereby, the antiproton results are closer to the measured
electron data than the results for proton impact since the former both share
the same charge \cite{anti:knud92}.

%In \cite{anti:luhr08a} it is shown that 
%Using the scaled hydrogen atom \hscal\ as model target for calculations of the
%total cross sections for ionization of a  \hmol\ molecule can give
%reasonable results~\cite{anti:luhr08a}.

%as has been seen in \cite{anti:luhr08a}. % due to the by
                                % construction correct ionization potential. 
%In contrast to the case of ionization as well as to the astonishing good
%agreement for excitation at high energies between the description with \vmod\
% %present model potential   
%and the measurements in Fig.~\ref{fig:cs_ex_H2}
%the excitation cross section for \hscal, however, clearly shows a different
%behavior than the experimental data. 

The \hscal\ results for protons and antiprotons also coincide for high impact
energies as expected in the first Born limit. 
However, the excitation cross sections for \hscal\ in Fig.~\ref{fig:cs_ex_H2}
as well as in~\cite{anti:luhr08a}
clearly show a different behavior than the experimental data and the results
obtained with the potential \vmod . 

In contrast to the present results, the use of \hscal\ as a target model in
calculations of total ionization cross sections of an \hmol\ molecule can
yield to a certain extent reasonable results~\cite{anti:luhr08a}. The mixed
capability of \hscal\ 
in describing the \hmol\ molecule may be explained in the following way. On
the one hand, concerning ionization, the ionization potential is in both
models, i.e.,  \vmod\ and   $V_{\rm scal}$, by definition correct. On
the other hand, the potentials differ in their $r$ 
dependence and the short-range as well as the long-range behavior of 
$V_{\rm scal}$ obviously disagrees with that of an \hmol\ molecule. This leads
to a poor description of the bound states and finally to wrong excitation cross
sections. This 
is in accordance with the deviations of the OS in
Fig.~\ref{fig:oscillator_strength_H2} which indicate too large excitation
probability for \hscal\ at $R=1.4$\au

%
%%%%%%%%%%%%%%%%%%%%%%%%%%%%%%%%%%%%%%% 
% 
% 
% 
\subsection{Helium atom} 
\label{sec:he_atom}
\begin{figure}[t] 
    \begin{center} 
      \includegraphics[width=0.48\textwidth]{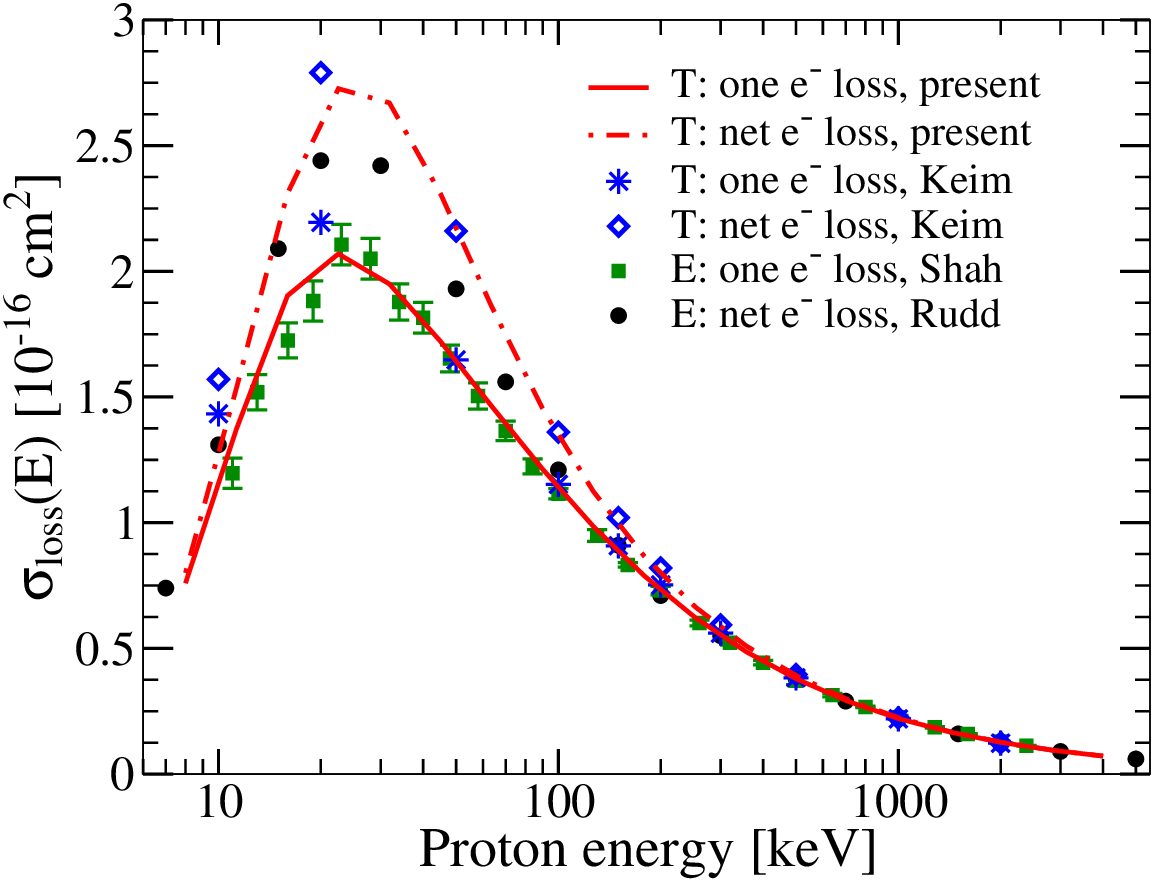} 
      \caption{(Color online) Electron loss cross section of He collisions
        with protons. 
        One-electron loss. Theory:
        red solid curve, \vmod\ with $\alpha=0.8791$;
        blue stars, Keim {\it et al.}~\cite{sct:keim05b}.
        Experiment:
        green squares, Shah and Gilbody~\cite{sct:shah85,sct:shah89}.
        Net electron loss. Theory:
        red dash--dotted curve, \vmod\ with $\alpha=0.8791$;
        blue diamonds, Keim {\it et al.}~\cite{sct:keim05b}.
        Experiment:
        black circles, Rudd {\it et al.}~\cite{sct:rudd83}.
        \label{fig:e_loss_he} } 
    \end{center} 
\end{figure} 
In the limit $R\rightarrow 0$ with $\alpha=0.8791$ the model potential \vmod\
can be used for the description of a helium atom. Obviously, various different 
one-electron potentials have already been proposed in order to describe He
atoms. There are, e.g., the well-known Thomas-Fermi and Hartree models as well
as the Hartree-Fock-Slater (HFS) model \cite{gen:slat51} which
includes a local exchange correction and was applied, e.g., in
\cite{sct:guly95}. Another approach is the optimized potential method (OPM)
discussed in \cite{sct:kirc97,gen:aash78}  which was used to calculated $p$
+ He loss cross sections in \cite{sct:keim05b}. 

In Fig.~\ref{fig:e_loss_he} electron loss cross sections (the sum of
ionization and capture) are shown for collisions of protons with He atoms. The
present results were obtained with the same method which was employed for the
$p$ and \pb\ collisions with \hmol\ in section \ref{sec:ex_cross_section} and
which was discussed in detail in \cite{anti:luhr08,anti:luhr08a}. Measurements
of the one-electron loss were performed by  Shah and
Gilbody~\cite{sct:shah85,sct:shah89}. Cross sections for the net 
electron loss were experimentally determined by Rudd {\it et
  al.}~\cite{sct:rudd83}. Calculations of the one-electron and net electron 
loss were done by Keim {\it et al.}~\cite{sct:keim05b} using the OPM with a
time-independent effective potential.   

In the present results for the net electron loss the probabilities for double
capture, double ionization, and transfer ionization are counted twice in order
to get the correct number of electrons lost in the collision process. All
theoretical data points for the net electron loss by Keim {\it et al.}
coincide fully with the present findings apart from those for the two lowest
impact energies (10 and 20\,keV) which are clearly higher than the present
ones. The present net loss cross sections reproduce the experimental data by
Rudd {\it et al.} to a great extent. However, in the energy range $20 < E<
100$\,keV the experimental data are smaller than the outcome of both
theoretical investigations.

In the case of the one-electron loss the present findings match the
experimental results by Shah and Gilbody well in the whole impact energy 
range of the protons.  Again all theoretical data points
by Keim {\it et al.} coincide fully with the present ones besides those for
the two lowest impact energies which have again larger values. 
Finally, it can be concluded that in addition to \hmol\ the proposed model
potential $V_{\rm mpd}$ is also capable of a simple description of He atoms
which is consistent with the OPM without response.

%  the potential curve of the present
% model \vmod\ for $R\rightarrow 0$ with $\alpha=0.8791$. Also given is the
% curve of the effective one-electron model potential proposed by Hartree in
% \cite{gen:hart57} for helium. It can be seen that the differences are only
% minute. The maximal relative deviation occurs at $r=1.05$\au\ and is
% 2.38\%. The $r$ range around the largest deviation is shown in the inset on an
% enlarged scale. 
% Due to the small differences between the potential curves also the bound state
% wave functions are practically the same for both models and the relative
% deviation of the bound state energies is at most 0.2\%. No additional
% comparisons of \vmod\ in the limit $R\rightarrow 0$ are presented since it
% does not vary substantially from the  He model potential which is well known
% and was frequently used. Recently it was employed, e.g., in
% \cite{sfm:peng08,sfm:chel04}.  

%  
%%%%%%%%%%%%%%%%%%%%%%%%%%%%%%%%%%%%%%%% 
% 
% 
%  

\section{Conclusion} 
 \label{sec:summary} 

A simple model potential \vmod\ for an effective one-electron, single-centered
description of the \hmol\ molecule has been proposed and its properties have
been examined. The potential is
unambiguously determined by the correct ionization potential of the \hmol\
molecule and allows for the description of \hmol\ at an arbitrary internuclear
distance $R$. Thereby, also the nuclear motion can be considered to a some
extent.  

The model potential was used for various applications in the range $0.8\le R
\le 2.5$\au\ \
The energetically-lowest, dipole-allowed excitation energies, transition
moments as well as oscillator strengths of the \hmol\ molecule are represented
well by the present model.  The model was furthermore employed for the
calculation of the 
photoionization cross section of \hmol\ and a partial excitation cross section
in collisions of \hmol\ with charged particles. In both applications
experimental and also theoretical literature data could be well described by 
the present results obtained with the model potential. 

Concerning the scaled hydrogen atom as model for \hmol\ the results for
ionization are satisfying while bound states properties and therefore also
excitation cross sections are not reproduced adequately.

The satisfying description of results for \hmol\ molecules justifies on the
one hand the choice of the ionization potential of \hmol\ as a criterion for
adjusting \vmod\ to a certain internuclear distance.
On the other hand, together with the surpassing simplicity of the model which
includes an approximate analytic expression for the ionization potential, it
suggests its applicability to a large number of further problems. To name only
some, 
there are, e.g., the description of the electronic structure of \hmol\
molecules in \hmol\ clusters, \hmol\ molecules interacting with external
fields or with particles, and finally also the modeling of He atoms in the
limit $R\rightarrow 0$.   

It can be concluded that the \hmol\ molecule is in many cases
surprisingly well described by a single-electron, one-center model. This
means, that in these cases the effects of charge anisotropy and two-electron
effects are small.

\begin{center} 
 {\bf ACKNOWLEDGMENTS}  
\end{center} 
The authors wish to thank H. J. L\"udde for a discussion on effective
one-electron models. The authors also would like to thank F. Mart{\'i}n and
T. Kirchner for providing cross sections in numerical form.
%as well as providing theoretical data.
The authors are grateful to BMBF (FLAIR Horizon), DFG, and {\it Stifterverband
  f\"ur die deutsche Wissenschaft} for financial support.

%%%%%%%%%%%%%%%%%%%%%%%%%%%%%%%%%%%%%%%%%%%%%%%%%%%%%%%%%%%%%% 
 
%\bibliography{anti,dia,gen,nu,sfm,sct} 

\end{document}